\documentclass[aps,amsmath,showpacs,showkeys,nofootinbib, 
superscriptaddress]{revtex4} 
\usepackage[dvips]{graphicx} 
 
\begin{document} 

\title{\textbf{Coupled channels calculation of a $\pi\Lambda N$ quasibound 
state}}

\author{H.~ Garcilazo} 
\email{humberto@esfm.ipn.mx} 
\affiliation{Escuela Superior de F\' \i sica y Matem\'aticas \\ 
Instituto Polit\'ecnico Nacional, Edificio 9, 
07738 M\'exico D.F., Mexico} 

\author{A.~Gal} 
\email{avragal@vms.huji.ac.il} 
\affiliation{Racah Institute of Physics, The Hebrew University, 
Jerusalem 91904, Israel} 

\date{\today} 

\begin{abstract} 
We extend the study of a $J^P=2^+,I=\frac{3}{2}$ $\pi\Lambda N$ quasibound 
state [Phys. Rev. D {\bf 78}, 014013 (2008)] by solving nonrelativistic 
Faddeev equations, using $^3S_1 - {^3}D_1~,~\Lambda N - \Sigma N$ coupled 
channels Chiral Quark Model local interactions, and $\pi N$ and coupled 
$\pi \Lambda - \pi \Sigma $ separable interactions fitted to the position 
and decay parameters of the $\Delta(1232)$ and $\Sigma(1385)$ resonances, 
respectively. The results exhibit a strong sensitivity to the $p$-wave 
pion-hyperon interaction, with a $\pi\Lambda N$ quasibound state persisting 
over a wide range of acceptable parametrizations. 
\end{abstract} 



\maketitle 


\section{Introduction} 

The success of the nonrelativistic quark model (QM) during the 1970s in 
reproducing the SU(3) octet and decuplet baryon masses in terms of $3q$ 
configurations was followed by QM studies of $6q$ configurations that 
aimed particularly at elucidating the baryon-baryon short-range dynamics 
and making related predictions for dibaryon bound or quasibound states. 
It is remarkable that decades of experimental searches for dibaryons have 
so far yielded no unambiguous evidence for a dibaryon state. 
In the nonstrange sector, where the quark cluster calculations for $L=0$ 
$6q$ configurations \cite{OYa80} suggest only a weakly bound $\Delta\Delta$ 
dibaryon with $(J^P,I)=(3^+,0)$, there is a recent indication from 
$np\to d\pi\pi$ reactions at CELSIUS-WASA for a resonance structure at 
$M_R \approx 2.36$ GeV and $\Gamma_R \approx 80$ MeV that might suggest 
a $\Delta\Delta$ dibaryon bound by about 100 MeV, but still about 200 MeV 
above the $d\pi\pi$ threshold \cite{BBB09}. In the strange sector, Jaffe's 
dibaryon $H$ \cite{Jaf77} with strangeness $S=-2$ and quantum numbers 
$(J^P,I)=(0^+,0)$ which was predicted as a genuinely bound state well below 
the $\Lambda\Lambda$ threshold, perhaps the most cited ever prediction made 
for any dibaryon, has not been confirmed experimentally to date in spite of 
several extensive searches \cite{Bas97}. Another equally ambitious early 
prediction was made by Goldman {\it et al.} \cite{Gold87}, also using 
a variant of the MIT bag model, for $(J^P;I)=(1^+,2^+;\frac{1}{2})$ $S=-3$ 
dibaryons dominated by $\Omega N$ structure and lying below the $\Xi\Lambda$ 
threshold. More realistic quark cluster calculations by Oka {\it et al.} 
\cite{OSI83}, applying resonating group techniques, did not confirm Jaffe's 
deeply bound $H$, placing it just below the $\Xi N$ threshold as a resonance 
about 26 MeV above the $\Lambda\Lambda$ threshold. The underlying binding 
mechanism common to all of these orbital angular momentum $L=0$ configurations 
is the dominance of the color-magnetic interaction for gluon exchange between 
quarks, a feature emphasized by Oka \cite{Oka88} who systematically studied 
$L=0$ dibaryon configurations that may benefit from a short-range attraction. 
Following earlier quark cluster calculations \cite{OYa80,OSI83}, these 
calculations resulted in no strange dibaryon bound states, and for the 
$\Omega N$-dominated $S=-3$ bound-state configurations predicted in 
Ref.~\cite{Gold87}, in particular, only a $(J^P,I)=(2^+,\frac{1}{2})$ 
quasibound state resulted. 

For strangeness $S=-1$, which is the focus of the present work, no $L=0$ 
dibaryons have been suggested for the lowest energy $I=\frac{1}{2}$ 
$\Lambda N - \Sigma N$ coupled channels, where the long-range pion exchange 
interaction is dominant, particularly for the $^3S_1-{^3}{D_1}$ system through 
the tensor component. Although old $K^-d\to\pi^-\Lambda p$ data \cite{Tan69} 
had suggested resonant $\Lambda p$ structures at the $\Sigma N$ threshold and 
10 MeV above it, a $(J^P,I)=(1^+,\frac{1}{2})$ $\Sigma N$ quasibound state 
is not necessarily required in order to reproduce the general shape of the 
$\Lambda p$ spectrum as shown by multichannel Faddeev calculations 
\cite{TGE79,TDD86}. Several low-lying $L=1$ $\Lambda N$ resonances were 
predicted in singlet and triplet configurations in a QM study 
by Mulders {\it et al.} \cite{MAS80}, but negative results, particularly 
for the singlet resonance, were reported in dedicated $K^-$ initiated 
experiments \cite{JHK92,CDG92} near the $\Sigma N$ threshold. At higher 
energies, Oka's analysis \cite{Oka88} drew attention to a $(J^P,I) = (2^+,
\frac{1}{2})$ dibaryon predominantly of a $\Sigma(1385)N - \Sigma\Delta(1232)$ 
coupled channels structure resonating about the $\Sigma\Delta(1232)$ 
threshold, approximately 100 MeV above the lower $\Sigma(1385)N$ threshold. 
We note that these two channels are substantially higher in mass, by about 
300 MeV, than the $S=-1$ thresholds of $\Lambda N$ and $\Sigma N$. 

In a recent paper \cite{GALGA} we studied within three-body Faddeev 
calculations the possible existence of a $\pi\Lambda N$ quasibound state, 
driven by the two-body $(^{2S+1}L_J,I)=(^2P_{\frac{3}{2}},\frac{3}{2})$ 
$\pi N$ resonance $\Delta(1232)$ and the $(^2P_{\frac{3}{2}},1)$ $\pi\Lambda$ 
resonance $\Sigma(1385)$, for a $\Lambda N$ $(^3S_1,\frac{1}{2})$ 
configuration, all of which were represented by means of single-channel 
separable potentials. The coupling to the pionless $\Sigma N$ channel, with 
threshold about 60 MeV below the $\pi\Lambda N$ threshold, was disregarded. 
It was felt that this coupling was mostly responsible for the width of the 
$\pi\Lambda N$ quasibound state. The three-body channel $(J^P,I)=(2^+,
\frac{3}{2})$ was selected since all the angular momenta, spins, and isospins 
in this channel have maximum values and, therefore, it is likely to benefit 
from maximal attraction of both $\Delta(1232)$ and $\Sigma(1385)$ resonances. 
This opportunity is unique to strange and charmed systems: a similar choice 
of $(J^P,I)=(2^+,2)$ for $\pi NN$, with each $\pi N$ pair interacting in 
the $(\frac{3}{2}^+,\frac{3}{2})$ resonating channel, implies a $(1^+,1)$ 
Pauli-forbidden $NN$ configuration. In terms of dibaryons, the $\pi\Lambda N$ 
$(J^P,I)=(2^+,\frac{3}{2})$ quasibound state is a deeply bound $\Sigma(1385)N 
- \Lambda\Delta(1232)$ $L=0$ dibaryon, at energy considerably below the 
$(J^P,I)=(2^+,\frac{1}{2})$ $\Sigma(1385)N - \Sigma\Delta(1232)$ $L=0$ 
dibaryon suggested by the quark cluster model of Ref.~\cite{Oka88}.

Whereas the interactions in the pion-baryon resonating channels in first 
approximation are adequately represented by rank-one attractive separable 
potentials, the baryon-baryon interaction requires a rank-two separable 
potential to simulate both the attraction and repulsion that meson-exchange 
models normally yield. Indeed, we found a strong dependence of the calculated 
$\pi\Lambda N$ binding energy on the balance between repulsion and attraction 
in the $^3S_1$ $\Lambda N$ channel \cite{GALGA}. It is therefore suggestive 
to consider a more realistic hyperon-nucleon interaction in the $J^P=1^+$ 
channel. In the present work we used the hyperon-nucleon ($YN$) Chiral Quark 
Model (CQM) interaction described in Refs.~\cite{SALA1,SALA2} in terms of 
$^3S_1 - {^3}D_1~,~\Lambda N - \Sigma N$ coupled channels {\it local} 
potentials. For consistency, we also generalized our previous single-channel 
model of $\Sigma(1385)$ as a $\pi\Lambda$ resonance to a family of 
pion-hyperon ($\pi Y$) interaction models, in terms of $\pi\Lambda -\pi\Sigma$ 
coupled channels separable interactions fitted to the position, width and 
decay branching ratios of $\Sigma(1385)$. Furthermore, we studied the 
dependence of the calculated $\pi\Lambda N$ binding energy on the $\pi Y$ 
interaction. 

In our previous work \cite{GALGA}, based on separable potentials, 
we considered both a nonrelativistic and a relativistic three-body 
formalism from which we deduced that the nonrelativistic results 
do not change much when the relativistic formalism is used instead. 
This is relevant for the validity of the results of the present 
work which are based on the hyperon-nucleon interaction derived 
from the CQM within a {\it nonrelativistic} formalism. Therefore, 
in the present calculation we consider only a nonrelativistic framework. 
The results of the present three-body Faddeev calculations leave wide 
room for the existence of a $(2^+,\frac{3}{2})$ $\pi\Lambda N$ quasibound 
state indicating, however, a strong dependence on the short-range behavior 
of the least known $\pi Y$ and $YN$ two-body subsystems. 

The plan of the paper is as follows. In Sec. II we describe the two-body 
interactions in the pion-nucleon, pion-hyperon and hyperon-nucleon subsystems. 
In Sec. III we derive the Faddeev equations of the pion-nucleon-hyperon 
system. Finally, we discuss our results in Sec. IV and summarize the work 
in Sec. V.

\section{The two-body subsystems} 

We will denote the hyperon, nucleon, and pion as particles 1, 2, and 3, 
respectively, and refer to the two-body subsystems by a subscript for 
the spectator particle. Thus, pion-nucleon is subsystem 1, pion-hyperon 
is subsystem 2, and hyperon-nucleon is subsystem 3. The conventional 
reduced masses are given by 
\begin{equation} 
\eta_i^\alpha=\frac{m_jm_k}{m_j+m_k},\,\,\,\,\,\,\,\,\,
\nu_i^\alpha=\frac{m_i(m_j+m_k)}{m_i+m_j+m_k},  
\label{eq0} 
\end{equation} 
where a superscript $\alpha=\Lambda,\Sigma$ has been added to indicate 
whether particle 1 is a $\Lambda$ or a $\Sigma$ hyperon and, obviously, 
\begin{equation} 
\eta_1^\Lambda \equiv \eta_1^\Sigma \equiv \eta_1=\frac{m_\pi m_N}{m_\pi+m_N}. 
\label{eq0pp} 
\end{equation} 
However, an average hyperon mass 
\begin{equation} 
m_Y=\frac{m_\Lambda+m_\Sigma}{2}  
\label{eq13} 
\end{equation} 
was used in the following reduced masses: 
\begin{equation} 
\nu_2=\frac{m_N(m_\pi+m_Y)}{m_N+m_\pi+m_Y},\,\,\,\,\,\,\,\,\,
\nu_3=\frac{m_\pi(m_N+m_Y)}{m_\pi+m_N+m_Y}. 
\label{eq12} 
\end{equation} 

The $\pi Y$ and $YN$ amplitudes are given by $2\times 2$ matrices, to account 
for the coupling between $\pi\Lambda$ and $\pi\Sigma$ and between $\Lambda N$ 
and $\Sigma N$, respectively. The $\pi N$ amplitude in the three-body system 
is also given by $2\times 2$ matrix, since the energy dependence of the 
two-body subsystem depends on whether the spectator particle is a $\Lambda$ 
or a $\Sigma$.

\subsection{The pion-nucleon subsystem} 

Since the $\pi N$ subsystem is dominated by the $\Delta$(1232) 
resonance, a rank-one separable interaction is considered sufficient:  
\begin{equation} 
<p_1|V_1|p_1^\prime>=\gamma_1g_1(p_1)g_1(p_1^\prime)~, 
\label{eq1} 
\end{equation} 
so that the corresponding two-body $t$-matrix is given by 
\begin{equation} 
<p_1|t_1(E)|p_1^\prime>=g_1(p_1)\tau_1(E)g_1(p_1^\prime)~, 
\label{eq2} 
\end{equation} 
where $E=p_0^2/2\eta_1$ with $p_0$ the correct 
relativistic $\pi N$ center of mass (c.m.) momentum and 
\begin{equation} 
\tau_1^{-1}(E)= 1/\gamma_1-\int_0^\infty p_1^2dp_1\frac{g_1^2(p_1)} 
{E-p_1^2/2\eta_1+i\epsilon}~. 
\label{eq3} 
\end{equation} 
The form factor $g_1(p_1)$ was obtained from a very good fit of the $P_{33}$ 
phase shift \cite{ARNDT} for $0\le T_{\rm lab}\le 250$ MeV in the form 
\begin{equation} 
g_1(p_1)=p_1[e^{-p_1^2/\beta_1^2}+A_1p_1^2e^{-p_1^2/\alpha_1^2}]~, 
\label{eq4p} 
\end{equation} 
with $\gamma_1=-0.03317$ fm$^4$, $A_1=0.2$ fm$^2$, $\beta_1=1.31$ fm$^{-1}$, 
and $\alpha_1=3.2112$ fm$^{-1}$. 

In the three-body calculation, when the $\pi N$ subsystem is embedded in the 
$\pi YN$ system, the energy argument of the isobar propagator $\tau_1(E)$ 
depends on whether the spectator hyperon is a $\Lambda$ or a $\Sigma$, 
so that the separable $\pi N$ ampitude (\ref{eq2}) takes the form 
\begin{equation} 
t_1=|g_1>\begin{pmatrix} 
\tau_\Lambda(q_1) & 0 \cr 
0 & \tau_\Sigma(q_1)  \cr\end{pmatrix} 
<g_1|, 
\label{eq4} 
\end{equation} 
where 
\begin{equation} 
\tau_\alpha(q_1) =
\tau_1(E-\delta_{\alpha\Sigma}\Delta E-q_1^2/2\nu_1^\alpha);\,\,\,\,\,\,\,\, 
\alpha=\Lambda,\Sigma, 
\label{eq5} 
\end{equation} 
with $q_1$ the relative momentum between the hyperon and the 
$\pi N$ subsystem and 
\begin{equation} 
\Delta E=m_\Sigma-m_\Lambda. 
\label{eq5p} 
\end{equation}

\subsection{The pion-hyperon subsystem} 

The $\pi Y$ subsystem is dominated by the $\Sigma$(1385) $p$-wave resonance 
which decays mainly into $\pi\Lambda$ and $\pi\Sigma$ with branching ratios 
of $(87.0 \pm 1.5)\%$ and $(11.7 \pm 1.5)\%$, respectively \cite{PDG}. 
To account for the coupling $\pi\Lambda - \pi\Sigma$, we assume a coupled 
channels separable interaction: 
\begin{equation} 
<p_2|V_2^{\alpha\beta}|p_2^\prime>=\gamma_2g_2^\alpha(p_2) 
g_2^\beta(p_2^\prime);\,\,\,\,\,\,\,\,\alpha,\beta=\Lambda,\Sigma, 
\label{eq6} 
\end{equation} 
so that the corresponding two-body $t$-matrix is given by 
\begin{equation} 
<p_2|t_2(E)|p_2^\prime>=g_2^\alpha(p_2)\tau_2(E)g_2^\beta(p_2^\prime); 
\,\,\,\,\,\,\,\,\alpha,\beta=\Lambda,\Sigma, 
\label{eq7} 
\end{equation} 
with 
\begin{equation} 
\tau_2^{-1}(E)= 1/\gamma_2-\int_0^\infty p_2^2dp_2\frac{[g_2^\Lambda(p_2)]^2} 
{E-p_2^2/2\eta_2^\Lambda+i\epsilon} 
-\int_0^\infty p_2^2dp_2\frac{[g_2^\Sigma(p_2)]^2} 
{E-\Delta E-p_2^2/2\eta_2^\Sigma+i\epsilon}. 
\label{eq8} 
\end{equation} 
Again, $E=p_0^2/2\eta_2^\Lambda$ where $p_0$ is the correct relativistic 
$\pi\Lambda$ c.m. momentum and $\Delta E$ is chosen such that the $\pi\Lambda$ 
momentum at the $\pi\Sigma$ threshold has its correct value, that is 
\begin{equation} 
\Delta E = \frac{[(m_\Sigma+m_\pi)^2-(m_\Lambda+m_\pi)^2][(m_\Sigma+m_\pi)^2 
-(m_\Lambda-m_\pi)^2]}{8\eta_2^\Lambda(m_\Sigma+m_\pi)^2}. 
\label{eq9} 
\end{equation} 
The $\pi Y$ $t$-matrix (\ref{eq7}) in the $\pi YN$ system may be written in 
compact notation as a $2\times 2$ matrix 
\begin{equation} 
t_2=\begin{pmatrix} |g_2^\Lambda> \cr |g_2^\Sigma> \cr\end{pmatrix} 
\tau_N(q_2) \begin{pmatrix} 
<g_2^\Lambda| & <g_2^\Sigma| \cr\end{pmatrix}, 
\label{eq10} 
\end{equation} 
where 
\begin{equation} 
\tau_N(q_2)=\tau_2(E-q_2^2/2\nu_2). 
\label{eq11} 
\end{equation} 

The form factors $g_2^Y(p_2)$ of the separable $\pi Y$ $p$-wave potentials 
were taken in the form 
\begin{equation} 
g_2^\Lambda(p_2)=p_2(1+A_2p_2^2)e^{-p_2^2/\alpha_2^2},\,\,\,\,\,\,\,\,\,
g_2^\Sigma(p_2)=c_2g_2^\Lambda(p_2). 
\label{eq14} 
\end{equation} 
Solutions exist for all values of $A_2$ between 0 and $\infty$. 
Therefore, in order to fit the position, width and decay branching 
ratios of $\Sigma(1385)$, we have at our disposal four free parameters: 
$A_2$, $\alpha_2$, $\gamma_2$ and $c_2$, which provide for varying one of 
these while adjusting the other three to the three pieces of data. We thus 
constructed five models (models A-E) by considering five values of the 
parameter $A_2$, as shown in Table \ref{tab1}. We also constructed a sixth 
model (model F) which shares the same range parameter $\alpha_2$ with model C 
but which neglects the coupling to the $\pi\Sigma$ channel ($c_2=0$), as was 
done in our previous calculation \cite{GALGA}. It is instructive to classify 
the various $\pi Y$ interaction form factors $g_2^Y(p_2)$ according to their 
root-mean-square (r.m.s.) momentum, using the following 
expression for the mean-square momentum $<p_2^2>_{g_2}$: 
\begin{equation} 
<p_2^2>_{g_2}~=~\frac{\int_0^\infty g_2(p_2)~p_2^2~d^3p_2}
{\int_0^\infty g_2(p_2)~d^3p_2}~=~3\alpha_2^2\frac{A_2\alpha_2^2+
\frac{1}{3}}{A_2\alpha_2^2+\frac{1}{2}}~\approx~3\alpha_2^2, 
\label{eq14.1} 
\end{equation} 
where the approximation owes to $2A_2\alpha_2^2>>1$. The resulting values 
of the r.m.s. momentum, listed in the last column of Table~\ref{tab1}, 
are close to $\sqrt{<p_2^2>_{g_2}}\approx 3.8~\rm{fm}^{-1}\approx 750$ MeV/c. 
For comparison, $\sqrt{<p_1^2>_{g_1}}= 5.55~\rm{fm}^{-1}\approx 1100$ MeV/c 
for the $\pi N$ form factor $g_1(p_1)$ of Eq.~(\ref{eq4p}).\footnote{This 
high-momentum value for $g_1$ does not rule out a spatial size of order 1 fm 
for $\Delta(1232)$. Indeed, if ${\tilde g}_1(r_1)$ is the Fourier transform 
of $g_1(p_1)$, for $\ell=1$, then $\sqrt{<r_1^2>_{{\tilde g}_1}}=0.875$ fm.} 
\begin{table} 
\caption{Five choices (A-E) of form factor parameters for the coupled channels 
$\pi Y$ subsystem, Eqs.~(\ref{eq6}) and (\ref{eq14}). The last line lists 
a single-channel $\pi \Lambda$ sixth model (F) with $c_2=0$. The last column 
lists values of the r.m.s. momentum [see Eq.~(\ref{eq14.1})].} 
\begin{ruledtabular} 
\begin{tabular}{cccccccc} 
& Model & $A_2~({\rm fm}^2)$ & $\alpha_2~({\rm fm}^{-1})$ & $\gamma_2~
({\rm fm}^4)$ & $c_2$ & $\sqrt{<p_2^2>_{g_2}}~({\rm fm}^{-1})$ & \\ 
\hline 
 &  A  &  0.8  & 2.41372   & 
   -0.00604931 & 0.890227  & 4.11 & \\ 
 &  B  &  1.0  & 2.29039   & 
   -0.00552272 & 0.925591  & 3.91 & \\ 
 &  C  &  1.2  & 2.20024   & 
   -0.00501334 & 0.956818  & 3.76 & \\ 
 &  D  &  1.5  & 2.10192   & 
   -0.00433208 & 0.997300  & 3.60 & \\ 
 &  E  &  1.8  & 2.03076   & 
   -0.00375829 & 1.03166   & 3.48 & \\ 
\hline 
 &  F  &  3.21 & 2.20024   & 
   -0.00149204 &   0       & 3.79 & \\ 
\end{tabular} 
\end{ruledtabular} 
\label{tab1} 
\end{table}

\subsection{The hyperon-nucleon subsystem} 

The $YN$ interaction derived from the chiral quark model is a local potential 
obtained by application of the Born-Oppenheimer approximation to the chiral 
quark-quark interaction (consisting of confinement, one-gluon exchange, 
pseudovector-meson exchange, and scalar-meson exchange) with a fully 
antisymmetrized six-quark wave function \cite{SALA1,SALA2,SALA3}. 
In the case of the $J^P=1^+$, $I=\frac{1}{2}$ channel, it leads to the 
following system of coupled equations: 
\begin{eqnarray} 
t_{\ell\ell^{\prime\prime}}^{\alpha\beta}(p_3,p_3^{\prime\prime};E)=&& 
V_{\ell\ell^{\prime\prime}}^{\alpha\beta}(p_3,p_3^{\prime\prime})+ 
\sum_{\gamma=\Lambda,\Sigma}\sum_{\ell^\prime=0,2}\int_0^\infty {p_3^\prime}^2 
d p_3^\prime
V_{\ell\ell^\prime}^{\alpha\gamma}(p_3,p_3^\prime) 
\nonumber \\ & & \times 
\frac{1}{E-\delta_{\gamma\Sigma}\Delta E -{p_3^\prime}^2/2\eta_3^\gamma 
+i\epsilon} 
t_{\ell^\prime\ell^{\prime\prime}}^{\gamma\beta} 
(p_3^\prime,p_3^{\prime\prime};E):\,\,\,\,\,\,\,\,\alpha,\beta=\Lambda,\Sigma, 
\label{eq15} 
\end{eqnarray} 
with $\alpha,\beta=\Lambda,\Sigma$, $\ell,\ell^{\prime\prime}=0,2$ and 
$E=p_0^2/2\eta_3^\Lambda$, where $p_0$ is the correct relativistic $\Lambda N$ 
c.m. momentum, and $\Delta E$ is chosen such that the $\Lambda N$ momentum at 
the $\Sigma N$ threshold has its correct value, that is 
\begin{equation} 
\Delta E = \frac{[(m_\Sigma+m_N)^2-(m_\Lambda+m_N)^2][(m_\Sigma+m_N)^2 
-(m_\Lambda-m_N)^2]}{8\eta_3^\Lambda(m_\Sigma+m_N)^2}. 
\label{eq16} 
\end{equation} 
The $YN$ $t$-matrix (\ref{eq15}) may be written in compact notation as 
a $2\times 2$ matrix 
\begin{equation} 
t_3=\begin{pmatrix} 
t^{\Lambda\Lambda} & t^{\Lambda\Sigma} \cr 
t^{\Sigma\Lambda} & t^{\Sigma\Sigma}  \cr\end{pmatrix}, 
\label{eq17} 
\end{equation} 
where each $YN$ $t$-matrix $t^{\alpha\beta}$ includes, in addition, 
a coupling between $S$ ($\ell=0$) and $D$ ($\ell=2$) waves.

\section{The three-body equations} 

The Faddeev equations for the bound-state problem 
\begin{equation} 
T_i=\sum_{j\ne i}t_iG_0T_j; \,\,\,\,\,\,\,\, i,j=1,2,3, 
\label{eq18} 
\end{equation} 
couple the amplitudes $T_1$, $T_2$ and $T_3$ together. Eliminating the 
amplitude $T_3$ in favor of $T_1$ and $T_2$, one obtains 
\begin{equation} 
T_1=t_1G_0t_3G_0T_1+(t_1+t_1G_0t_3)G_0T_2, 
\label{eq19} 
\end{equation} 
\begin{equation} 
T_2=t_2G_0t_3G_0T_2+(t_2+t_2G_0t_3)G_0T_1, 
\label{eq20} 
\end{equation} 
where, in order to allow for the $Y=(\Lambda,\Sigma)$ specification, one has 
\begin{equation} 
G_0=\begin{pmatrix} 
G_0^\Lambda & 0 \cr 
0 & G_0^\Sigma  \cr\end{pmatrix}, 
\label{eq20p} 
\end{equation} 
Since the two-body amplitudes $t_1$ and $t_2$ are separable 
[see Eqs.~(\ref{eq4}) and (\ref{eq10})], the three-body amplitudes 
$T_1$ and $T_2$ are of the form 
\begin{equation} 
T_1=|g_1>\begin{pmatrix} X_\Lambda \cr X_\Sigma \cr\end{pmatrix}, 
\label{eq21} 
\end{equation} 
\begin{equation} 
T_2=\begin{pmatrix} |g_2^\Lambda> \cr |g_2^\Sigma> \cr\end{pmatrix} X_N, 
\label{eq22} 
\end{equation} 
where the subscript of the amplitude $X$ indicates which particle is 
the spectator. Substitution of (\ref{eq21}) and (\ref{eq22}) into 
(\ref{eq19}) and 
(\ref{eq20}) leads to 
\begin{eqnarray} 
\begin{pmatrix} X_\Lambda \cr X_\Sigma \cr\end{pmatrix}=&& 
\begin{pmatrix}
\tau_\Lambda & 0 \cr 
0 & \tau_\Sigma  \cr\end{pmatrix} 
<g_1| 
\begin{pmatrix} 
G_0^\Lambda t^{\Lambda\Lambda}G_0^\Lambda 
& G_0^\Lambda t^{\Lambda\Sigma}G_0^\Sigma \cr 
G_0^\Sigma t^{\Sigma\Lambda}G_0^\Lambda 
& G_0^\Sigma t^{\Sigma\Sigma}G_0^\Sigma \cr\end{pmatrix}|g_1> 
\begin{pmatrix} X_\Lambda \cr X_\Sigma \cr\end{pmatrix} 
\nonumber \\ & & 
\nonumber \\ & & + 
\begin{pmatrix} 
\tau_\Lambda & 0 \cr 
0 & \tau_\Sigma  \cr\end{pmatrix} 
<g_1| 
\begin{pmatrix} 
G_0^\Lambda + 
G_0^\Lambda t^{\Lambda\Lambda}G_0^\Lambda  
& G_0^\Lambda t^{\Lambda\Sigma}G_0^\Sigma \cr 
G_0^\Sigma t^{\Sigma\Lambda}G_0^\Lambda & G_0^\Sigma 
+ G_0^\Sigma t^{\Sigma\Sigma}G_0^\Sigma  \cr\end{pmatrix} 
\begin{pmatrix} |g_2^\Lambda> \cr |g_2^\Sigma> \cr\end{pmatrix}X_N, 
\label{eq23} 
\end{eqnarray} 
\begin{eqnarray} 
X_N=& & 
\tau_N \begin{pmatrix} 
<g_2^\Lambda| & <g_2^\Sigma| \cr\end{pmatrix} 
\begin{pmatrix} 
G_0^\Lambda + 
G_0^\Lambda t^{\Lambda\Lambda}G_0^\Lambda 
& G_0^\Lambda t^{\Lambda\Sigma}G_0^\Sigma \cr 
G_0^\Sigma t^{\Sigma\Lambda}G_0^\Lambda & G_0^\Sigma 
+ G_0^\Sigma t^{\Sigma\Sigma}G_0^\Sigma  \cr\end{pmatrix}|g_1> 
\begin{pmatrix} X_\Lambda \cr X_\Sigma \cr\end{pmatrix} 
\nonumber \\ & & + 
\tau_N \begin{pmatrix} 
<g_2^\Lambda| & <g_2^\Sigma| \cr\end{pmatrix} 
\begin{pmatrix} 
G_0^\Lambda t^{\Lambda\Lambda}G_0^\Lambda 
& G_0^\Lambda t^{\Lambda\Sigma}G_0^\Sigma \cr 
G_0^\Sigma t^{\Sigma\Lambda}G_0^\Lambda 
& G_0^\Sigma t^{\Sigma\Sigma}G_0^\Sigma \cr\end{pmatrix} 
\begin{pmatrix} |g_2^\Lambda> \cr |g_2^\Sigma> \cr\end{pmatrix}X_N, 
\label{eq24} 
\end{eqnarray} 
which are integral equations in one continuous variable given explicitly by 
\begin{eqnarray} 
X_\alpha(q_1)=&& \tau_\alpha(q_1)\sum_{\beta=\Lambda,\Sigma} 
\int_0^\infty {q_1^\prime}^2 dq_1^\prime K^{\alpha\beta}(q_1,q_1^\prime) 
X_\beta(q_1^\prime) 
\nonumber \\ & & + 
\tau_\alpha(q_1) 
\int_0^\infty q_2^2 dq_2 K^{\alpha N}(q_1,q_2)X_N(q_2); 
\,\,\,\,\,\,\,\, \alpha=\Lambda,\Sigma, 
\label{eq25} 
\end{eqnarray} 
\begin{eqnarray} 
X_N(q_2)=&& \tau_N(q_2)\sum_{\alpha=\Lambda,\Sigma} 
\int_0^\infty q_1^2 dq_1 K^{N \alpha}(q_2,q_1) 
X_\alpha(q_1) 
\nonumber \\ & & + 
 \tau_N(q_2) 
\int_0^\infty {q_2^\prime}^2 dq_2^\prime K^{N N}(q_2,q_2^\prime) 
X_N(q_2^\prime). 
\label{eq26} 
\end{eqnarray} 
The kernels of these integral equations are given in the Appendix.

\section{Results and discussion} 

\begin{table} 
\caption{Binding energy of $\pi\Lambda N$ (in MeV) for five $\pi Y$ 
interaction models (A-E of Table~\ref{tab1}) and six CQM versions of the 
$^3S_1 - {^3}D_1$ $YN$ interaction fitted to given $\Lambda N$ scattering 
length $a$ and effective range $r_0$ (both in fm). 
The momentum $p_{\rm lab}(\delta=0)$ is the $\Lambda$ laboratory momentum 
(in MeV/c) where the $^3S_1$ $\Lambda N$ phase shift changes sign. 
The last line corresponds to switching off the $YN$ interaction.} 
\begin{ruledtabular} 
\begin{tabular}{cccccccccc} 
& $a$ & $r_0$ &  $p_{\rm lab}(\delta=0)$  & Model A 
& Model B  & Model C & Model D & Model E & \\ 
\hline 
&  -1.35 &   3.39  &   987  &   147  &   99  & 65  & 30  & 6 & \\ 
&  -1.40 &   3.32  &  1011  &   147  &   99  & 66  & 30  & 6 & \\ 
&  -1.64 &   3.09  &  1146  &   150  &  102  & 68  & 32  & 8 & \\ 
&  -1.71 &   3.03  &  1198  &   150  &  102  & 68  & 33  & 9 & \\ 
&  -1.78 &   2.98  &  1272  &   151  &  103  & 69  & 33  & 9 & \\ 
&  -1.86 &   2.93  &  1446  &   152  &  104  & 69  & 34  & 10 & \\ 
&   --   &   --   &    --   &   170  &  120  & 84  & 47  & 21 & \\  
\end{tabular} 
\end{ruledtabular} 
\label{tab2} 
\end{table} 

We applied the formalism of the previous section, using six different versions 
of the $YN$ interaction obtained from the CQM, all of which reproduce equally 
well the experimental low-energy $YN$ data \cite{SALA1,SALA2}. Results are 
listed in Table~\ref{tab2} from where it is clear that the $\pi\Lambda N$ 
binding energies are substantial for $\pi Y$ models with $A_2<1~{\rm fm}^2$. 
Generally, the higher the r.m.s. momentum of the $\pi Y$ form factor $g_2$, 
the stronger is the binding, as demonstrated in the table. Irrespective of 
which $\pi Y$ model is chosen, the $YN$ interaction always produces repulsion, 
thus lowering the calculated binding energy, as demonstrated by the results 
listed in the last line which corresponds to switching off the $YN$ 
interaction. This repulsive $YN$ effect owes its origin to the high-momentum 
components of the $\pi B$ form factors which within the three-body calculation 
highlight the short-range repulsive region of the $YN$ interaction. 

To demonstrate the model dependence of the three-body calculation within 
a given $\pi Y$ model, we assembled in Table~\ref{tab3} several binding 
energy results based on model C and also on its limitation to the 
$\pi \Lambda$ channel (model F listed in Table~\ref{tab1}). The $YN$ models 
included in this table invariably give $a=-1.40$~fm, whether limited to the 
$\Lambda N$ $^3S_1$ single channel or extended to the $^3S_1 - {{^3}D}_1$ 
$\Lambda N-\Sigma N$ coupled channels. Comparing the first two entries to each 
other, we conclude that the extension from a single $\Lambda N$ channel to 
$\Lambda N-\Sigma N$ coupled channels has very little effect (about 3 MeV 
additional attraction) within the $\pi \Lambda$ model F. In contrast, for the 
full $\pi Y$ model C, as in the last two entries, the extension from 
$\Lambda N$ to $YN$ models has a somewhat larger effect (about 7 MeV 
repulsion) and in the opposite direction. Within the $\pi Y$ model C, the full 
coupled channels $YN$ interaction contributes 18 MeV repulsion (third and 
fifth entries in Table~\ref{tab3}) to the three-body binding energy. Similar 
results hold for all other $\pi Y$ models. 

\begin{table} 
\caption{Comparison of $\pi\Lambda N$ binding energies (in MeV) calculated 
within the $\pi Y$ model C and its $\pi \Lambda$ limit model F (see 
Table~\ref{tab1}) for $YN$ coupled channels and $\Lambda N$ single-channel 
models with $a=-1.40$~fm, and for no $YN$ interaction.} 
\begin{ruledtabular} 
\begin{tabular}{ccccccc} 
& $\pi\Lambda$, $\Lambda N$ & $\pi\Lambda$, $YN$ & $\pi Y$, no $YN$ & 
$\pi Y$, $\Lambda N$ & $\pi Y$, $YN$ & \\ 
\hline 
& 93 & 96 & 84 & 73 & 66 & \\ 
\end{tabular}  
\end{ruledtabular} 
\label{tab3} 
\end{table} 

To discuss the model dependence of the three-body calculation within 
a given $YN$ CQM, we follow Ref.~\cite{GALGA} in singling out 
$p_{\rm lab}(\delta=0)$, the momentum where the $\Lambda N$ $^3S_1$ phase 
shift changes sign from attraction outside to repulsion inside, as a measure 
of the repulsive $YN$ effect. We notice in Table~\ref{tab2} that the CQM 
values of $p_{\rm lab}(\delta=0)$ are considerably larger than those obtained 
by other models \cite{RSY99,RYa06,POLIN}, signifying less repulsion in the 
CQM. In order to test whether the apparent lack of repulsion in the CQM $YN$ 
interaction is responsible for the large binding energies obtained for 
$A_2<1~{\rm fm}^2$, we added to the CQM with $a=-1.40$ fm and $r_0=3.32$ fm 
a short-range potential in the $^3S_1$ $\Lambda N$ partial wave of the form 
\begin{equation} 
V(r)=\gamma_R \frac{e^{-\beta_R r}}{r} - \gamma_A \frac{e^{-\beta_A r}}{r}, 
\label{eq50} 
\end{equation} 
with $\beta_R=10$ fm$^{-1}$ and $\gamma_R \ge 1000$ MeV fm, while the 
attractive term was adjusted to maintain the $\Lambda N$ scattering length 
$a=-1.35$ fm and the effective range $r_0$ as close as possible to 3.39 fm, 
so that the $YN$ observables are not changed noticeably. The overall effect 
of $V(r)$ is repulsive, as demonstrated in Table~\ref{tab4} for the $\pi Y$ 
model A, with $A_2=0.8~{\rm fm}^2$, where it is clearly seen that increase in 
the strength of the repulsive term lowers the value of $p_{\rm lab}(\delta=0)$ 
as well as lowering the $\pi\Lambda N$ binding energy. 

\begin{table} 
\caption{Binding energy of $\pi\Lambda N$ (in MeV) for the $\pi Y$ model A 
($A_2=0.8$ fm$^2$) and the CQM $YN$ interaction plus a short-range 
$\Lambda N$ potential $V(r)$, Eq.~(\ref{eq50}), with scattering length 
$a=-1.40$ fm and effective range $r_0=3.32$ fm. The strength parameters 
$\gamma$ are in units of MeV fm, the inverse range parameter $\beta_A$ is 
in units of fm$^{-1}$, $\beta_R=10$~fm$^{-1}$, and $p_{\rm lab}(\delta=0)$ 
is the laboratory momentum (in MeV/c) where the $^3S_1$ $\Lambda N$ phase 
shift changes sign.} 
\begin{ruledtabular} 
\begin{tabular}{ccccccc} 
& $\gamma_R$ & $\gamma_A$ & $\beta_A$ & $p_{\rm lab}(\delta=0)$ & $B(A_2=0.8)$ 
&  \\ 
\hline 
&  1000 &    240  & 5.371  &   873  &   107   & \\ 
&  2000 &    530  & 5.811  &   846  &    88   & \\ 
&  3000 &    720  & 5.749  &   822  &    70   & \\ 
&  4000 &    990  & 5.928  &   810  &    59   & \\ 
&  5000 &   1260  & 6.056  &   802  &    51   & \\ 
&  6000 &   1360  & 5.921  &   788  &    39   & \\ 
&  7000 &   1670  & 6.086  &   775  &    34   & \\ 
\end{tabular}  
\end{ruledtabular} 
\label{tab4} 
\end{table}

\section{Summary} 

In this work, we have extended the Faddeev equations study of 
a $(J^P,I)=(2^+,\frac{3}{2})$ quasibound $\pi\Lambda N$ state \cite{GALGA} 
from a $^3S_1$ $\Lambda N$ single-channel to 
$^3S_1 - {^3}D_1~,~\Lambda N - \Sigma N$ coupled channels, and from 
a $\pi \Lambda$ single-channel description of $\Sigma(1385)$ to 
$\pi \Lambda - \pi \Sigma$ coupled channels description. 
Local interaction potentials given by the CQM were used in the $YN$ sector, 
whereas one-rank separable potentials were used in the $\pi B$ sectors. 
We have shown within a nonrelativistic version of the Faddeev equations, 
but using semirelativistic kinematics, that the $\pi\Lambda N$ system 
is bound under a wide choice of parametrizations of the $\pi Y$ interaction 
form factor. The form factors of the $\pi B$ subsystems are sufficiently 
short ranged such that the pion undergoes almost coherently attraction to 
both baryons. The short-ranged repulsion between the two baryons in the 
CQM is insufficient to overcome the attraction gained by the pion unless 
the CQM is modified arbitrarily at very short distances to do this job. 
Altogether, the acceptable model dependence of the $\pi Y$ interaction 
form factor, and the uncertainty of the short-range behavior of the $YN$ 
interaction, leave plenty of room, theoretically, for a quasibound $S=-1$, 
$(J^P,I)=(2^+,\frac{3}{2})$, $\pi\Lambda N$ dibaryon. 

Before closing we list several production reactions, where the first two were 
already discussed in our previous paper \cite{GALGA}, in which to search for 
this $S=-1$ dibaryon here denoted $\cal D$: 
\begin{equation} 
K^- + d \to  {\cal D}^- + \pi^+~, \,\,\,\,\,\,\,\,\,\, 
\pi^- + d \to {\cal D}^- + K^+~,  
\label{eqKpiK}
\end{equation}   
\begin{equation} 
p + p \to {\cal D}^+ + K^+~.  
\label{eqpp} 
\end{equation} 
Correlated with the missing mass spectrum of the $\cal D$ dibaryon, for 
a forward outgoing meson, one should look for $\Sigma N$ decays that 
can be assigned to a $\Sigma N$ resonance with invariant mass $M_{\cal D}$. 
Total cross sections for the associated strangeness production 
$pp\to\Sigma NK^+$ near the hyperon production threshold have been reported 
from Juelich, for $\Sigma^0 p$ by the COSY-11 Collaboration \cite{cosy2004}, 
for $\Sigma^+ n$, also by COSY-11 \cite{cosy2006}, and by the ANKE 
Collaboration \cite{anke2007} and the HIRES Collaboration \cite{hires2010}, 
with conflicting results among all these $\Sigma^+ n$ reports. 
Old DISTO data for the reaction $pp\to\Lambda pK^+$ have been analysed to 
search for an intermediate $K^-pp$ quasibound state, with the astounding 
report of a broad resonance at the $\pi\Sigma N$ threshold \cite{disto2010}. 
Of course, this $I=\frac{1}{2}$ resonance cannot be assigned to 
a $I=\frac{3}{2}$ $\pi\Lambda N$ quasibound state, but forthcoming data from 
the FOPI detector Collaboration at GSI \cite{suzuki2009} could be analysed 
also with respect to a $\Sigma N$ rather than a $\Lambda p$ final state.

\section*{Acknowledgments} 
The research of H.M. was supported in part by COFAA-IPN (M\'exico) and that 
of A.G. by the EU Initiative FP7, HadronPhysics2, under Project 227431.

\section*{Appendix: Expressions for the kernels of the integral equations 
Eqs.~(\ref{eq25}) and (\ref{eq26})} 

We provide here detailed expressions for the kernels appearing in 
the integral equations Eqs.~(\ref{eq25}) and (\ref{eq26}). 

\begin{eqnarray}  
K^{\alpha\beta}(q_1,q_1^\prime)=&& \frac{1}{4}\sum_{\ell,\ell^\prime=0,2} 
\int_0^\infty q_3^2 dq_3 \int_{-1}^1 d{\rm cos}\theta\int_{-1}^1 
d {\rm cos}\theta^\prime 
\nonumber \\ & & \times 
 g_1(p_1^\alpha)G_0^\alpha(p_1^\alpha,q_1)b_{13}^\alpha 
A_{13,1\alpha}^{10\ell 1}(q_1,q_3,{\rm cos}\theta) 
\nonumber \\ & & \times 
t_{\ell\ell^\prime}^{\alpha\beta}(p_3^\alpha,{p^\prime}_3^\beta;
E-q_3^2/2\nu_3)b_{31}^\beta A_{31,1\beta}^{\ell^\prime 1 1 0}(q_3,q_1^\prime,
{\rm cos}\theta^\prime)G_0^\beta({p^\prime}_1^\beta,q_1^\prime)
g_1({p^\prime}_1^\beta), 
\label{eq27} 
\end{eqnarray} 

\begin{eqnarray} 
K^{\alpha N}(q_1,q_2)=&&\frac{1}{2}\int_{-1}^1 d{\rm cos}\theta 
g_1(p_1^\alpha)G_0^\alpha(p_1^\alpha,q_1)b_{12}^\alpha A_{12,1\alpha}^{1010}
(q_1,q_2,{\rm cos}\theta) 
g_2^\alpha(p_2^\alpha) 
\nonumber \\ & & + 
 \frac{1}{4}\sum_{\ell,\ell^\prime=0,2} 
\sum_{\beta=\Lambda,\Sigma} 
\int_0^\infty q_3^2 dq_3 \int_{-1}^1 d{\rm cos}\theta\int_{-1}^1 
d {\rm cos}\theta^\prime 
\nonumber \\ & & \times 
 g_1(p_1^\alpha)G_0^\alpha(p_1^\alpha,q_1)b_{13}^\alpha 
A_{13,1\alpha}^{10\ell 1}(q_1,q_3,{\rm cos}\theta) 
\nonumber \\ & & \times 
t_{\ell\ell^\prime}^{\alpha\beta}(p_3^\alpha,{p^\prime}_3^\beta;E-q_3^2/2\nu_3)
b_{32}^\beta A_{32,1\beta}^{\ell^\prime 1 1 0}(q_3,q_2,{\rm cos}\theta^\prime) 
G_0^\beta(p_2^\beta,q_2)g_2^\beta(p_2^\beta), 
\label{eq34} 
\end{eqnarray} 

\begin{equation} 
K^{N \alpha}(q_2,q_1)=K^{\alpha N}(q_1,q_2), 
\label{eq222} 
\end{equation} 

\begin{eqnarray} 
K^{N N}(q_2,q_2^\prime)=&& \frac{1}{4}\sum_{\ell,\ell^\prime=0,2} 
\sum_{\alpha,\beta=\Lambda,\Sigma} 
\int_0^\infty q_3^2 dq_3 \int_{-1}^1 d{\rm cos}\theta\int_{-1}^1 
d {\rm cos}\theta^\prime 
\nonumber \\ & & \times 
 g_2^\alpha(p_2^\alpha)G_0^\alpha(p_2^\alpha,q_2)b_{23}^\alpha 
A_{23,1\alpha}^{10\ell 1}(q_2,q_3,{\rm cos}\theta) 
\nonumber \\ & & \times 
t_{\ell\ell^\prime}^{\alpha\beta}(p_3^\alpha,{p^\prime}_3^\beta;E-q_3^2/2\nu_3)
b_{32}^\beta A_{32,1\beta}^{\ell^\prime 1 1 0}(q_3,q_2^\prime,{\rm cos}
\theta^\prime)G_0^\beta({p^\prime}_2^\beta,q_2^\prime)g_2^\beta
({p^\prime}_2^\beta), 
\label{eq39} 
\end{eqnarray} 
with 

\begin{equation} 
G_0^\alpha(p_i,q_i)=\frac{1}{E -\delta_{\alpha\Sigma}\Delta E 
-p_i^2/2\eta_i^\alpha-q_i^2/2\nu_i^\alpha+i\epsilon}, 
\,\,\,\,\,\,\,\, \alpha=\Lambda,\Sigma. 
\label{eq40} 
\end{equation} 

The orbital angular momentum recoupling coefficients 
$A_{ij,L\alpha}^{\ell_i\lambda_i\ell_j\lambda_j}(q_i,q_j,{\rm cos}\theta)= 
A_{ji,L\alpha}^{\ell_j\lambda_j\ell_i\lambda_i}(q_j,q_i,{\rm cos}\theta)$, 
and isospin recoupling coefficients $b_{ij}^\alpha=b_{ji}^\alpha$, are 
calculated by consideration of a cyclic pair $ij$. (The spin recoupling 
coefficients are all equal to $1$.) For isospin we have 

\begin{equation} 
b_{ij}^\alpha = 
(-)^{i_j+\tau_j-I}\sqrt{(2i_i+1)(2i_j+1)} \, W(\tau_j\tau_kI\tau_i;i_ii_j), 
\label{eq41} 
\end{equation} 
where $W$ is a Racah coefficient. 
If $\alpha=\Lambda$ then  $\tau_1=0$, $\tau_2=\frac{1}{2}$, 
$\tau_3=1$, $i_1=\frac{3}{2}$, 
$i_2=1$, $i_3=\frac{1}{2}$, and $I=\frac{3}{2}$ so that 
$b_{12}^\Lambda=b_{31}^\Lambda=b_{23}^\Lambda=1$. If $\alpha=\Sigma$ 
we have instead $\tau_1=1$ so that $b_{12}^\Sigma=\sqrt{5/6}$, 
$b_{31}^\Sigma=-{\sqrt{5}}/3$, and $b_{23}^\Sigma=-1/\sqrt{6}$. 

The orbital angular momentum recoupling coefficients are given by 

\begin{eqnarray} 
A_{ij,L\alpha}^{\ell_i\lambda_i\ell_j\lambda_j}(q_i,q_j,{\rm cos}\theta)=&& 
\frac{1}{2L+1}\sum_{M m_i m_j}C^{\ell_i \lambda_i L}_{m_i,M-m_i,M} 
C^{\ell_j \lambda_j L}_{m_j,M-m_j,M}\Gamma_{\ell_i m_i}\Gamma_{\lambda_i 
M-m_i} 
\nonumber \\ && \times 
\Gamma_{\ell_j m_j} 
\Gamma_{\lambda_j M-m_j} 
{\rm cos}(-M\theta-m_i\theta_i^\alpha 
+m_j\theta_j^\alpha);\,\,\,\,\,\,\,\,\,\alpha=\Lambda,\Sigma, 
\label{eq42} 
\end{eqnarray} 
where $\Gamma_{\ell m}=0$ for odd values of $\ell -m$, and 

\begin{equation} 
\Gamma_{\ell m}=\frac{(-1)^{(\ell+m)/2} 
\sqrt{(2\ell+1)(\ell+m)!(\ell-m)!} 
}{2^\ell[(\ell+m)/2]![(\ell-m)/2]!}, 
\label{eq43} 
\end{equation} 
for even values of $\ell-m$. The angles $\theta_i^\alpha$ and 
$\theta_j^\alpha$ are obtained from 

\begin{equation} 
{\rm cos}\theta_i^\alpha=-\frac{q_j{\rm cos}\theta+q_i a_{ij}^\alpha}
{p_i^\alpha}, 
\label{eq44} 
\end{equation} 

\begin{equation} 
{\rm cos}\theta_j^\alpha=\frac{q_i{\rm cos}\theta+q_j a_{ji}^\alpha}{p_j^
\alpha}, 
\label{eq45} 
\end{equation} 

\begin{equation} 
p_i^\alpha=\sqrt{q_j^2+(q_i a_{ij}^\alpha)^2+2q_iq_j a_{ij}^\alpha {\rm cos}
\theta}, 
\label{eq46} 
\end{equation} 

\begin{equation} 
p_j^\alpha=\sqrt{q_i^2+(q_j a_{ji}^\alpha)^2+2q_iq_j a_{ji}^\alpha {\rm cos}
\theta}, 
\label{eq47} 
\end{equation} 
where 

\begin{equation} 
a_{ij}^\alpha=\frac{m_j}{m_j+m_k},\,\,\,\,\,\,\,\,\,
a_{ji}^\alpha=\frac{m_i}{m_i+m_k},  
\label{eq48} 
\end{equation} 
with $m_1=m_\alpha$; $\alpha=\Lambda,\Sigma$. 
Equations (\ref{eq46}) and (\ref{eq47}) provide also the relative momenta 
appearing in Eqs. (\ref{eq27})-(\ref{eq39}).

\end{document}